# Brown Dwarfs in the Pleiades Cluster Confirmed by the Lithium Test [1]

R. Rebolo, E. L. Martín
Instituto de Astrofísica de Canarias, E-38200 La Laguna, Tenerife. Spain

G. Basri, G. W. Marcy
Department of Astronomy, University of California, Berkeley, Berkeley, CA 94720. USA

and

M. R. Zapatero Osorio
Instituto de Astrofísica de Canarias, E-38200 La Laguna, Tenerife. Spain

e-mail addresses: rrl@iac.es, ege@iac.es, basri@soleil.berkeley.edu
gmarcy@etoile.berkeley.edu, mosorio@iac.es

## ABSTRACT

We present 10 m Keck spectra of the two Pleiades brown dwarfs Teide 1 and Calar 3 showing a clear detection of the $\lambda$ 670.8 nm Li resonance line. In Teide 1, we have also obtained evidence for the presence of the subordinate line at $\lambda$ 812.6 nm. A high Li abundance (log $N$(Li)$\geq$2.5), consistent with little if any depletion, is inferred from the observed lines. Since Pleiades brown dwarfs are unable to burn Li the significant preservation of this fragile element confirms the substellar nature of our two objects. Regardless of their age, their low luminosities and Li content place Teide 1 and Calar 3 comfortably in the genuine brown dwarf realm. Given the probable age of the Pleiades cluster, their masses are estimated at 55$\pm$15 Jupiter masses.

[1] Based on observations obtained at the W.M. Keck Observatory, which is operated jointly by the University of California and the Californian Institute of Technology.

## 1. Introduction

In stellar interiors $^7$Li nuclei are destroyed via proton collisions at relatively low temperatures ($\sim 2\times10^6$ K). This element has long been used as a tracer (see e.g. Rebolo 1991; Michaud & Charbonneau 1991) of internal structure in stars of different types. The strong convection of very low-mass stars ($M \leq 0.3\ M_\odot$) causes an extremely efficient mixing of Li and indeed significant depletion has been observed in these stars, even in the very young ones (Zapatero Osorio et al. 1996a; García López, Rebolo, & Martín 1994; Martín, Rebolo, & Magazzù 1994). Objects with masses $M \leq 0.065\ M_\odot$ ($\sim 65\ M_{\rm J}$, Jupiter mass units) are well below the minimum mass for hydrogen burning and cannot reach the Li burning temperature so, unlike very low mass stars, they must preserve a significant amount of their initial Li content during their lifetime. This can be detected spectroscopically and therefore can provide a diagnostic of substellar nature for brown dwarf candidates (Rebolo, Martín, & Magazzù 1992; Magazzù, Martín, & Rebolo 1993).

At the age of the Pleiades cluster even the most massive brown dwarfs (80–60 $M_{\rm J}$) should have preserved a large amount of their initial Li, but searches (Martín et al. 1994; Marcy, Basri, & Graham 1994) of the faintest proper motion members failed to detect it. An obvious conclusion was that brown dwarfs had not been discovered yet in the cluster. This prompted the search for new fainter and cooler candidates that led to the discovery of Teide 1 and Calar 3 (Rebolo, Zapatero Osorio, & Martín 1995; Zapatero Osorio, Rebolo, & Martín 1996b). These are two very late-type dwarfs near the center of the cluster with radial velocities, photometric and spectroscopic properties and a proper motion measurement in the case of Teide 1 consistent with Pleiades membership (Rebolo et al. 1995; Zapatero Osorio et al. 1996b; Martín, Rebolo, & Zapatero Osorio 1996). Their luminosities and effective temperatures qualify them, according to all the available evolutionary models, as brown dwarfs. However, a definitive confirmation of their substellar nature can only be obtained via the Li test. Such confirmation is a key step to consolidate our understanding of brown dwarf interiors. Very recently, Li has been discovered (Basri, Marcy, & Graham 1996) in PPl 15, a photometric member (Stauffer, Hamilton, & Probst 1994) of the Pleiades about one magnitude brighter than Teide 1 and Calar 3, which strongly encouraged us to extend the search for Li in these new objects. PPl 15 appears to define the Li reappearance boundary in the Pleiades, and sits on the frontier between stars and brown dwarfs.

## 2. Observations and results

We report spectroscopic observations of Teide 1 and Calar 3 that provide clear evidence for the presence of Li in both. The observations were carried out at the 10 m Keck telescope (Mauna Kea, Hawaii) on 1995 November 19 and 20 and 1996 February 16 using the Low Resolution Imaging Spectrograph (LRIS). The 1200 g/mm grating provided a dispersion of 0.64 Å/pixel on the TEK (2048×2048 pix) detector in the region of the $\lambda$ 670.8 nm Li I resonance doublet and 0.6 Å/pixel in the region of the subordinate $\lambda$ 812.6 nm Li I line. The 1 arcsec slit width finally gave an effective resolution of $\sim$2 Å. The total integration time for Teide 1 in the spectral regions of the $\lambda$ 670.8 nm and $\lambda$ 812.6 nm Li I lines was 8.2 and 4 hr, respectively (individual exposures ranged from 60 to 80 min). Only one 60 min spectrum could be obtained for Calar 3 on 1996 February 16 at the wavelength of the resonance line due to bad weather.

The individual images have been independently processed using standard procedures within the IRAF[2] package, which included debias, flat field, optimal extraction and wavelength calibration using arc lamps. The spectra in the region of the Li resonance line were flux calibrated using the standard star G191–B2B (Massey et al. 1988). The final spectra of Teide 1 and Calar 3 centered at $\lambda$720 nm are shown in Fig. 1. H$\alpha$ is seen in emission (equivalent widths of 4.5$\pm$0.5 Å and 6.5$\pm$0.5 Å, respectively) and lines of K I at $\lambda$ 766.49, 769.90 nm, as well as TiO and VO bandheads are clearly detected. The new, more accurate radial velocities of Teide 1 and Calar 3 are 5.2$\pm$2.0 and 1$\pm$8 km s$^{-1}$, fully consistent with Pleiades membership and previous measurements (Rebolo et al. 1995; Martín et al. 1996). They were determined by cross-correlating with the spectrum of vB 10 ($v_{\rm r}$ = 35 km/s, Goldberg 1995) obtained with the same instrumental configuration. The $\lambda$ 750 nm VO absorption index (Kirkpatrick, Henry, & Simons 1995), previously employed for spectral classification (Rebolo et al. 1995), was also measured in the new spectra resulting in values of 1.144$\pm$0.001 and 1.130$\pm$0.002 in Teide 1 and Calar 3 respectively, in very good agreement with measurements for other very late-type dwarfs.

The Li I line at $\lambda$ 670.8 nm is clearly present in the spectra of Teide 1 and Calar 3 with remarkable strength as compared with the nearby TiO bands as can be seen in Fig. 2(a). An accurate measurement of the equivalent width of the Li I $\lambda$ 670.8 nm line cannot be performed at our spectral resolution because of the strong blending with the nearby TiO lines. In order to determine approximate equivalent widths, we subtracted the spectrum of other very late-type dwarfs (vB 10, LHS 2065) observed with the same configuration, for which much higher resolution spectra do not show any evidence of Li absorption, and then measured the equivalent width of the feature at the position of the resonance doublet. We found these "pseudo-equivalent widths" to be 1.0$\pm$0.2 Å and 1.8$\pm$0.4 Å in Teide 1 and Calar 3, respectively. They are useful as a reference for future searches although they should be used with caution to derive an abundance from the available theoretical curves of growth (Pavlenko et al. 1995) since these curves were computed considering pure Li absorption. The same procedure was applied to the final spectrum of Teide 1 in the region of the Li I at $\lambda$ 812.6 nm. After subtraction of the spectrum of a typical M8 dwarf (obtained averaging the spectra of LP412–27 and vB 10), an absorption is apparent (see Fig. 2(b)) at this precise wavelength with a pseudo-equivalent width of $\sim$ 100$\pm$25 mÅ, which we interpret as additional evidence for the presence of Li in the atmosphere. It would be desirable in principle to confirm this feature with higher spectral resolution and S/N ratio.

[2]IRAF is distributed by National Optical Astronomy Observatories, which is operated by the Association of Universities for Research in Astronomy, Inc., under contract with the National Science Foundation.

## 3. Lithium abundances and brown dwarf status

To estimate the Li abundance in our objects we have used spectral synthesis computations based on Allard & Hauschildt (1995) (AH) model atmospheres for very cool dwarfs. The effective temperatures of Teide 1 and Calar 3 were estimated from a new spectral type classification using pseudo-continuum indices (Martín et al. 1996) and the $T_{\rm eff}$ calibration given in Kirkpatrick et al. (1993) which was revised (Kirkpatrick 1995) according to the new AH models. For M8 dwarfs this calibration yields an effective temperature of 2600$\pm$150 K. The M6.5 dwarf PPl 15 would have a $T_{\rm eff}$ about 200 K hotter. The Li abundance was determined by comparison with the available spectral synthesis (Pavlenko et al. 1995) and with new computations kindly provided by Ya. Pavlenko using various molecular line lists (Kurucz 1992). A set of spectra with several Li abundances were generated for the AH models with effective temperatures of 2500 and 2700 K. While these models describe reasonably well the TiO bands in the vicinity of the Li resonance line, cooler AH models produce too broad and deep TiO absorption bands which do not fit the observed spectrum and appear to indicate that the effective temperature of our two objects is indeed higher than 2500 K. Unfortunately, the limitations in the available molecular line lists and the uncertainties in the effective temperature prevent a very precise determination of the Li abundance. We should note that there are other effective temperature calibrations in the literature (Brett 1995; Tsuji, Ohnaka, & Aoki 1996) which give lower values by 300–600 K for very late-type dwarfs. Several computations have been conducted to determine the sensitivity of the Li resonance doublet to uncertainties in the effective temperature. It can be safely concluded that Teide 1 and Calar 3 have a rather high Li abundance in the range log $N$(Li) = 2.5–3.3 (where the usual scale is log $N$(H) = 12). At present we cannot claim any difference in the Li content between Calar 3 and Teide 1. Our preliminary analysis of the Li I $\lambda$ 812.6 nm line in Teide 1 gives Li abundances a factor of 2–3 higher than the resonance line, strongly suggesting that Li has been significantly preserved in our objects. It is known that the Li resonance line is sensitive to chromospheric effects (Pavlenko et al.

1995; Houdebine & Doyle 1995) and might underestimate the true atmospheric abundance in late-type dwarfs. We intend to pursue this question in a subsequent investigation.

The use of Li abundance and luminosity provides a very efficient way to constrain masses at the very bottom of the Pleiades sequence (Basri et al. 1996). We can determine the luminosity of Teide 1 and Calar 3 using optical and infrared observations (Zapatero Osorio et al. 1996b, 1996c) which give $I$ = 18.8, 18.7 and $K$ = 15.1, 14.9, respectively. Using the calibration of bolometric magnitude versus ($I$–$K$) color (Tinney, Mould, & Reid 1993), the resulting values are log $L/L_\odot$ = –3.18±0.10 and −3.11±0.10, for Teide 1 and Calar 3. The value for Teide 1 is, in fact, a factor 2 lower than the luminosity estimated from the spectral type and $I$ magnitude in Rebolo et al. (1995). The new spectral classification as M8 (Martín et al. 1996) gives a luminosity of –3.18 for Teide 1 and −3.16 for Calar 3 (Monet et al. 1992), which are consistent with those obtained from the ($I$–$K$) color. For PPl 15 we adopted $I$ = 17.8, $K$ = 14.1 (Stauffer et al. 1994; Zapatero Osorio et al. 1996c) and a spectral type M6.5 (Martín et al. 1996). Its luminosity, –2.80±0.10, was obtained as the average of those resulting from the previous two calibrations. This value is consistent with that obtained by Basri et al. (1996) and implies that PPl 15 has a mass of about 80±10 $M_J$ and a most likely age of ∼115–120 Myr. Using various sets of computations, the Li abundance and luminosity constrain the masses of Teide 1 and Calar 3 to be below 70 $M_J$, no matter what their age (Nelson, Rappaport, & Chiang 1993; Chabrier, Baraffe, & Plez 1996; D'Antona 1996). Hence, they are confidently below the hydrogen burning mass limit. Given the plausible age interval for the cluster (70–140 Myr), the mass range is 70–40 $M_J$. If the age of PPl 15 were indeed representative of the very low mass objects in the cluster, we infer a likely mass of 55 $M_J$ for Teide 1 and Calar 3. In Table 1 we summarize the most relevant parameters of our objects.

## 4. The faint end of Li depletion in the Pleiades

In Fig. 3 we compare the Li abundances of Teide 1 and Calar 3 with those of Pleiades members in a wide mass interval. F and early G-type stars (effective temperatures above 6000 K) show that the Pleiades stars were formed with an initial abundance about log $N$(Li) = 3.1±0.15 (Boesgaard, Budge, & Ramsay 1988). Pre-main sequence convective mixing causes the depletion of Li in late G, K and early/mid M-dwarfs (effective temperatures 5500–3000 K) (García López et al. 1994; Martín et al. 1994; Marcy et al. 1994; Soderblom et al. 1993). The first indication of the expected reappearance of Li at the borderline between stars and brown dwarfs is provided by PPl 15 at an effective temperature slightly below 3000 K. This object has retained about 10% of its initial Li (Basri et al. 1996). At cooler effective temperatures, corresponding to lower masses, Teide 1 and Calar 3 clearly confirm the expectations of a Li abundance higher than that of PPl 15.

The origin of the remarkable spread in Li abundances in Pleiades stars with effective temperatures in the range 5500–4000 K has been subject of discussion for several years. At least two interpretations have been considered so far, one of which proposes the existence of a large age dispersion within the cluster members (Duncan & Jones 1983), while the other suggests that rotation influences the Li depletion of individual stars (see Martín & Claret 1996 and references therein). It is well known that the Li depletion around the substellar limit ($T_{\rm eff} < 3000$ K) is extremely sensitive to age (Stringfellow 1989; Magazzù et al. 1993; Nelson et al. 1993; Chabrier et al. 1996). The possible role of rotation in the Li depletion of these objects has to be investigated both from the theoretical and observational sides to take advantage of the reappearance of this element to shed light on the problem.

## 5. Final remarks

We have inferred a mass of about 55 $M_{\rm J}$ for Teide 1 and Calar 3 from their Li abundances and luminosities. They are the first genuine brown dwarfs in the Pleiades and hence it is interesting to use them to discuss the cluster mass function beyond the substellar limit. Our objects were discovered in an $R$, $I$ survey of the central region of the Pleiades cluster (Zapatero Osorio et al. 1996b). Teide 1 is located 15' away from the cluster center while Calar 3 is at $\sim$1° southeast from the core. The total area covered by the survey was 578 arcmin$^2$, only about 1% of the cluster. There is no reason to think that any of the studied regions is anomalous regarding their substellar contents. If Pleiades brown dwarfs have a spatial distribution similar to early and mid-M dwarfs, the expected number of objects like Teide 1 and Calar 3 in the whole cluster is in the range 100–200 as previously discussed in Rebolo et al. (1995). The cluster mass function appears to increase beyond the substellar limit, but it seems that no reasonable extrapolation of the mass function would allow brown dwarfs to dominate the mass of the cluster.

In this work we have shown that the two best brown dwarf candidates in the photometric survey of Zapatero Osorio et al. (1996b) have succesfully passed the Li test. In the future, objects with similar or fainter photometric magnitudes, and with radial velocity and proper motion that guarantee membership in the Pleiades should be considered as *bona fide* brown dwarfs. It would not be necessary to apply the Li test to all of them. Nevertheless, since the evolutionary models predict a high sensitivity of the Li abundance to age for masses in the range 80–60 $M_{\rm J}$, we consider it interesting to continue the observations of Li beyond the substellar limit to probe a possible age spread in the cluster.

*Acknowledgments*: We thank Ya. Pavlenko, F. D'Antona, I. Baraffe and G. Chabrier for providing theoretical computations and for useful discussions. Financial support for E.L.M. and R.R. participation in the Keck observations was provided by the Spanish DGICYT project no. PB92–0434–C02. Support for G.W.M. was provided by NASA Grant NAGW 3182 .

Table 1. Pleiades least massive objects.

| | PPl 15 | Calar 3 | Teide 1 |
|---|---|---|---|
| $I$ | 17.82[a] | 18.73[b] | 18.80[b] |
| $(I-K)$[c] | 3.66 | 3.79 | 3.69 |
| SpT[d] | M6.5 | M8 | M8 |
| $v_{\rm r}$ (km/s) | 4.6±4.0[e] | 1.0±8.0 | 5.2±2.0 |
| VO | 1.06±0.05[d] | 1.130±0.002 | 1.144±0.001 |
| $EW_{H\alpha}$ (Å) | 5.5[e], 11.5[d] | 6.5±1.0 | 4.5±1.0 |
| $EW_{Li_{\lambda 670.8}}$ (Å) | 0.5[e] | 1.8±0.4 | 1.0±0.2 |
| $EW_{Li_{\lambda 812.6}}$ (mÅ) | ... | ... | 100±25 |
| log $N$(Li) | 1.16[e] | 2.5–3.3 | 2.5–3.3 |
| log $(L/L_{\odot})$ | –2.80±0.10 | –3.11±0.10 | –3.18±0.10 |
| $T_{\rm eff}$ (K) | 2800±150 | 2600±150 | 2600±150 |
| Mass ($M_{\rm J}$) | 80±10 | 55±15 | 55±15 |

[a]Stauffer et al. (1994).

[b]Zapatero Osorio et al. (1996b).

[c]Zapatero Osorio et al. (1996c).

[d]Martín et al. (1996).

[e]Basri et al. (1996).

Fig. 1.— The optical spectrum of Teide 1 (above) and Calar 3 (below) showing very similar features. Positions of several atomic lines and molecular bands are indicated. An offset of 7 has been added to the spectrum of Teide 1 for clarity.

Fig. 2.— (a) The spectral region of the Li I $\lambda$ 670.8 nm line in Teide 1 and Calar 3. Overplotted to Teide 1 is the spectrum of vB 10 (dotted line). Also plotted for comparison is the spectrum (Basri et al. 1996) of PPl 15 degraded to the same resolution than the previous ones. (b) The spectral region of the Li I $\lambda$ 812.6 nm line in Teide 1 (solid line) and overplotted (dotted line) the average spectrum of the two M8 dwarfs LP 412-31 and vB 10. The substraction of these two spectra is also plotted. The two strong absorption features are Na I lines at $\lambda$ 818.33 and 819.48 nm.

Fig. 3.— Li abundance versus effective temperature for Pleiades members in a wide mass range. Abundances are given in the usual scale log $N$(H) = 12. Open circles and arrows (which indicate upper limits) are taken from the literature (Boesgaard et al. 1988; Soderblom et al. 1993; García López et al. 1994; Martín et al. 1994; Marcy et al. 1994). The Li abundance of PPl 15 has been taken from Basri et al. (1996).

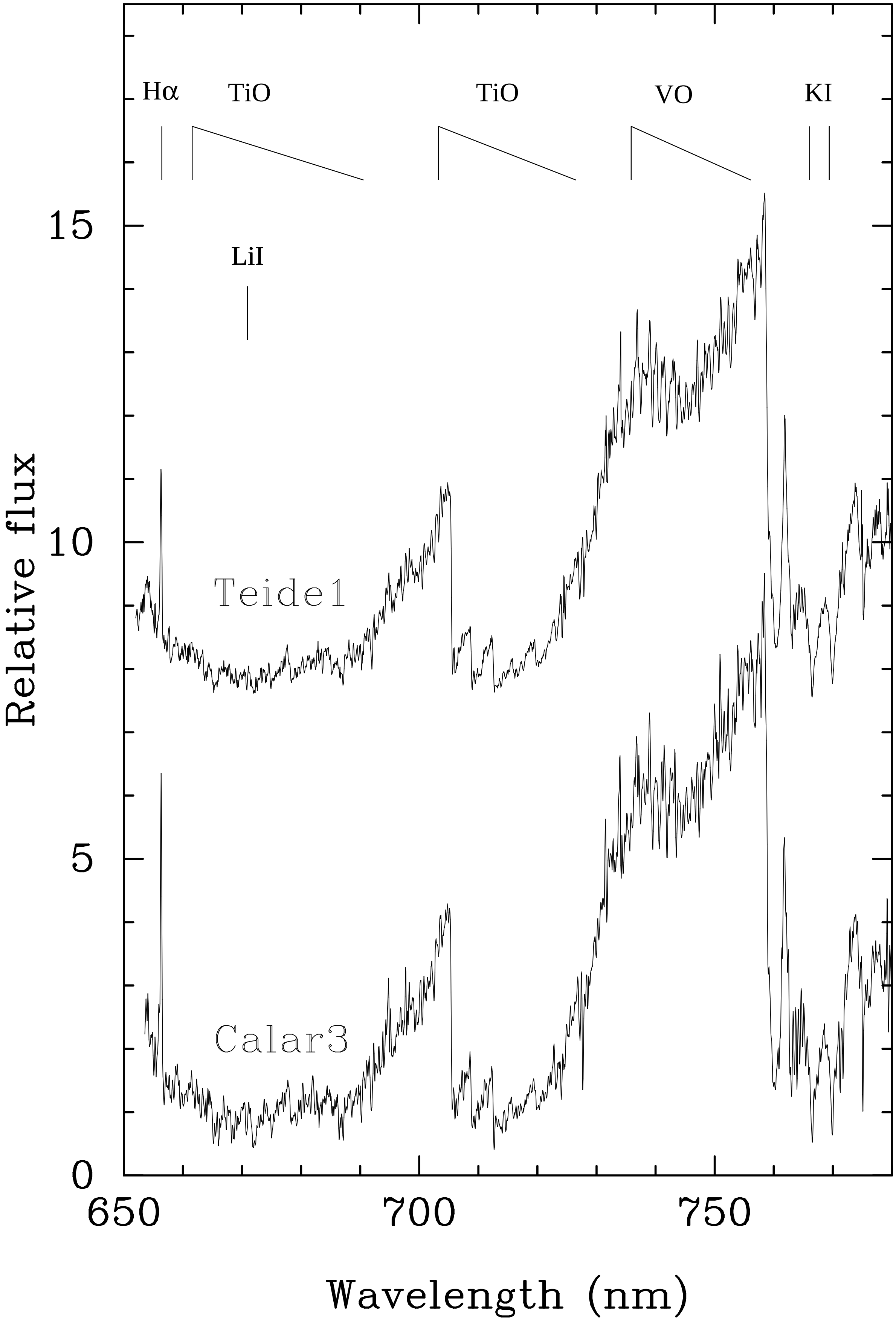

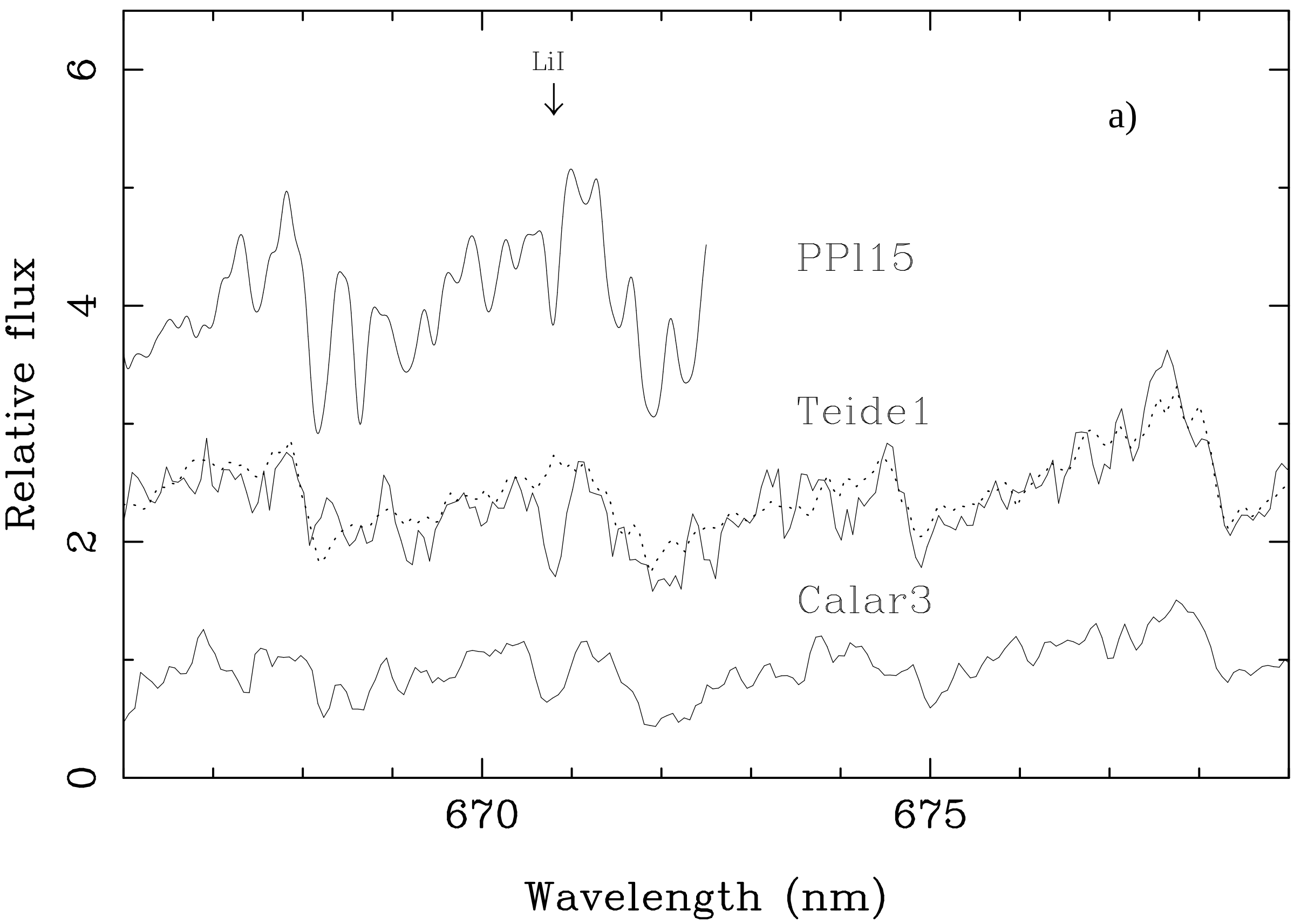

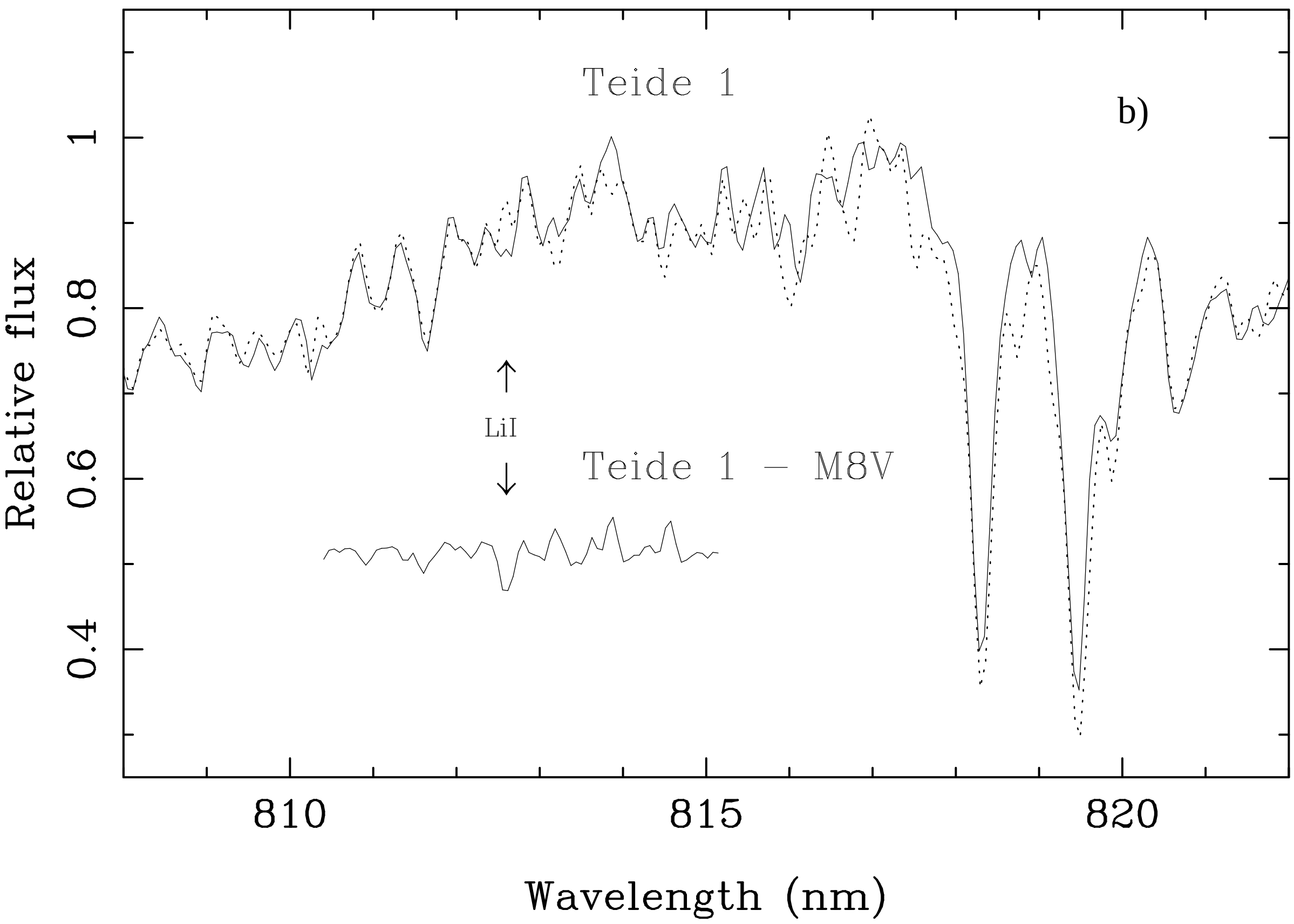

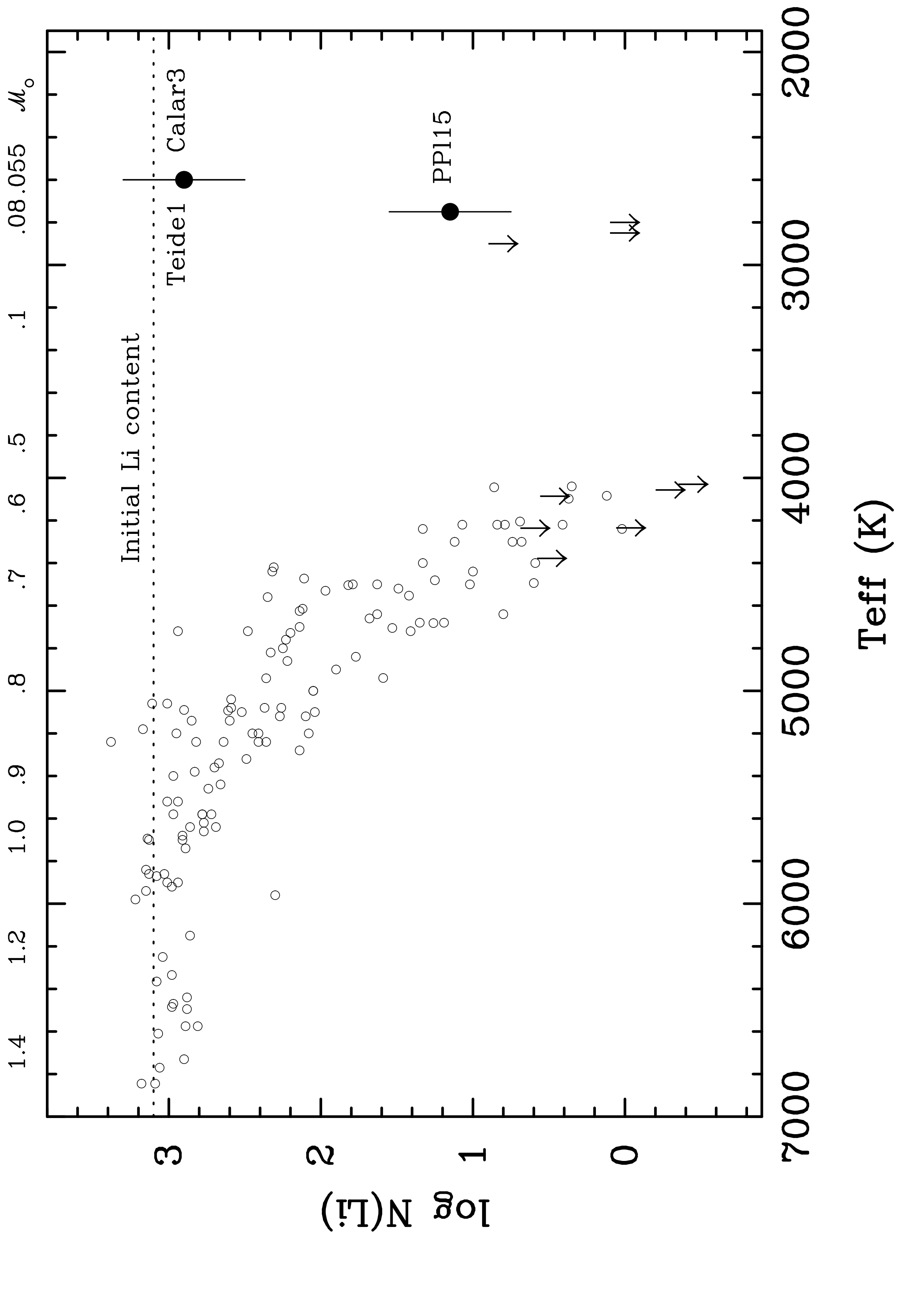
Teff (K)
log N(Li)
Initial Li content
Teide1
Calar3
PP115
1.4 1.2 1.0 .9 .8 .7 .6 .5 .1 .08.055 M⊙
7000 6000 5000 4000 3000 2000
0 1 2 3